\documentclass[10pt,letterpaper,twocolumn,english,final,floatfix,showpacs,showkeys,citesort,preprintnumbers,prd,nofootinbib,superscriptaddress]{revtex4-1}
\usepackage[T1]{fontenc}
\usepackage[latin1]{inputenc}
\usepackage{color}
\usepackage{babel}
\usepackage{float}
\usepackage{calc}
\usepackage{amsmath}
\usepackage{amssymb}
\usepackage{graphicx}
\usepackage[unicode=true,pdfusetitle,
 bookmarks=true,bookmarksnumbered=false,bookmarksopen=false,
 breaklinks=false,pdfborder={0 0 1},colorlinks=false]
 {hyperref}

\usepackage{paralist}

\makeatletter

\pdfpageheight\paperheight
\pdfpagewidth\paperwidth

 \@ifundefined{textcolor}{}
 {%
   \definecolor{BLACK}{gray}{0}
   \definecolor{WHITE}{gray}{1}
   \definecolor{RED}{rgb}{1,0,0}
   \definecolor{GREEN}{rgb}{0,1,0}
   \definecolor{BLUE}{rgb}{0,0,1}
   \definecolor{CYAN}{cmyk}{1,0,0,0}
   \definecolor{MAGENTA}{cmyk}{0,1,0,0}
   \definecolor{YELLOW}{cmyk}{0,0,1,0}
 }

\@ifundefined{definecolor}
 {\usepackage{color}}{}
\usepackage{babel}

\newcommand{\be}{\begin{equation}}
\newcommand{\ee}{\end{equation}}
\newcommand{\ba}{\begin{eqnarray}}
\newcommand{\ea}{\end{eqnarray}}

\@ifundefined{showcaptionsetup}{}{%
 \PassOptionsToPackage{caption=false}{subfig}}
\usepackage{subfig}
\makeatother

\begin{document} 

\preprint{
\vbox{
\null \vspace{0.3in}
\hbox{MITP/15-027}
\hbox{LPSC-15-082}
\hbox{SLAC-PUB-16258}
}
}

\title{\null \vspace{0.5in}
A review of the intrinsic heavy quark content of the nucleon}

\author{S. J.~Brodsky}
\thanks{sjbth@slac.stanford.edu}
\affiliation{SLAC National Accelerator Laboratory, Stanford University, Stanford,
	CA 94301, USA}

\author{A.~Kusina}
\thanks{kusina@lpsc.in2p3.fr}
\affiliation{Laboratoire de Physique Subatomique et de Cosmologie, Universit\'e Grenoble-Alpes, CNRS/IN2P3,
                   53 avenue des Martyrs, F-38026 Grenoble, France}

\author{F.~Lyonnet}
\thanks{flyonnet@smu.edu}
\affiliation{Southern Methodist University, Dallas, Texas 75275, USA}

\author{I.~Schienbein}
\thanks{ingo.schienbein@lpsc.in2p3.fr}
\affiliation{Laboratoire de Physique Subatomique et de Cosmologie, Universit\'e Grenoble-Alpes, CNRS/IN2P3,
                   53 avenue des Martyrs, F-38026 Grenoble, France}

\author{H.~Spiesberger}
\thanks{spiesber@uni-mainz.de}
\affiliation{PRISMA Cluster of Excellence, Institut f\"ur Physik, Johannes Gutenberg-Universit\"at, 55099 Mainz, Germany,
and Centre for Theoretical and Mathematical Physics and 
  Department of Physics, University of Cape Town, Rondebosch 7700, South Africa}

\author{R.~Vogt}
\thanks{vogt2@llnl.gov}
\affiliation{
Nuclear and Chemical Sciences Division, Lawrence Livermore National Laboratory, 
Livermore, CA 94551, USA \break
Physics Department, University of California at Davis, Davis, CA 95616, 
USA
}


%\pacs{{change 12.38.-t,13.15.+g,13.60.-r,24.85.+p}

\begin{abstract}
We present a review of the state-of-the-art of our understanding
of the intrinsic charm and bottom content of the nucleon.
We discuss theoretical calculations, constraints from global analyses,
and collider observables sensitive to the intrinsic heavy quark distributions.
A particular emphasis is put on the potential of 
a high-energy and high-luminosity fixed
target experiment using the LHC beams (AFTER@LHC) to search for
intrinsic charm.
\end{abstract}

\maketitle
\tableofcontents{}

\newpage{}

%%%%%%%%%%%%%%%%%%%%%%%%%%%%%%%%%%%%%%%%%%%%%%%%%%%%%%%%%%%%
%MAIN PART 
%%%%%%%%%%%%%%%%%%%%%%%%%%%%%%%%%%%%%%%%%%%%%%%%%%%%%%%%%%%%

\section{Introduction}
\label{sec:intro}
The existence of a nonperturbative intrinsic heavy quark component in the 
nucleon is a rigorous prediction of Quantum Chromodynamics (QCD). 
An unambiguous experimental confirmation is still missing and would 
represent a major discovery.  The goal of this article is to summarize our 
current understanding of this subject with a particular focus on
the potential of a high energy and high luminosity fixed-target experiment 
using the LHC beams (AFTER@LHC) 
\cite{Brodsky:2012vg,Lansberg:2012kf,Lansberg:2013wpx,Rakotozafindrabe:2013cmt} 
to search for intrinsic charm.

\begin{figure}
 \begin{center}
\includegraphics[width=7cm]{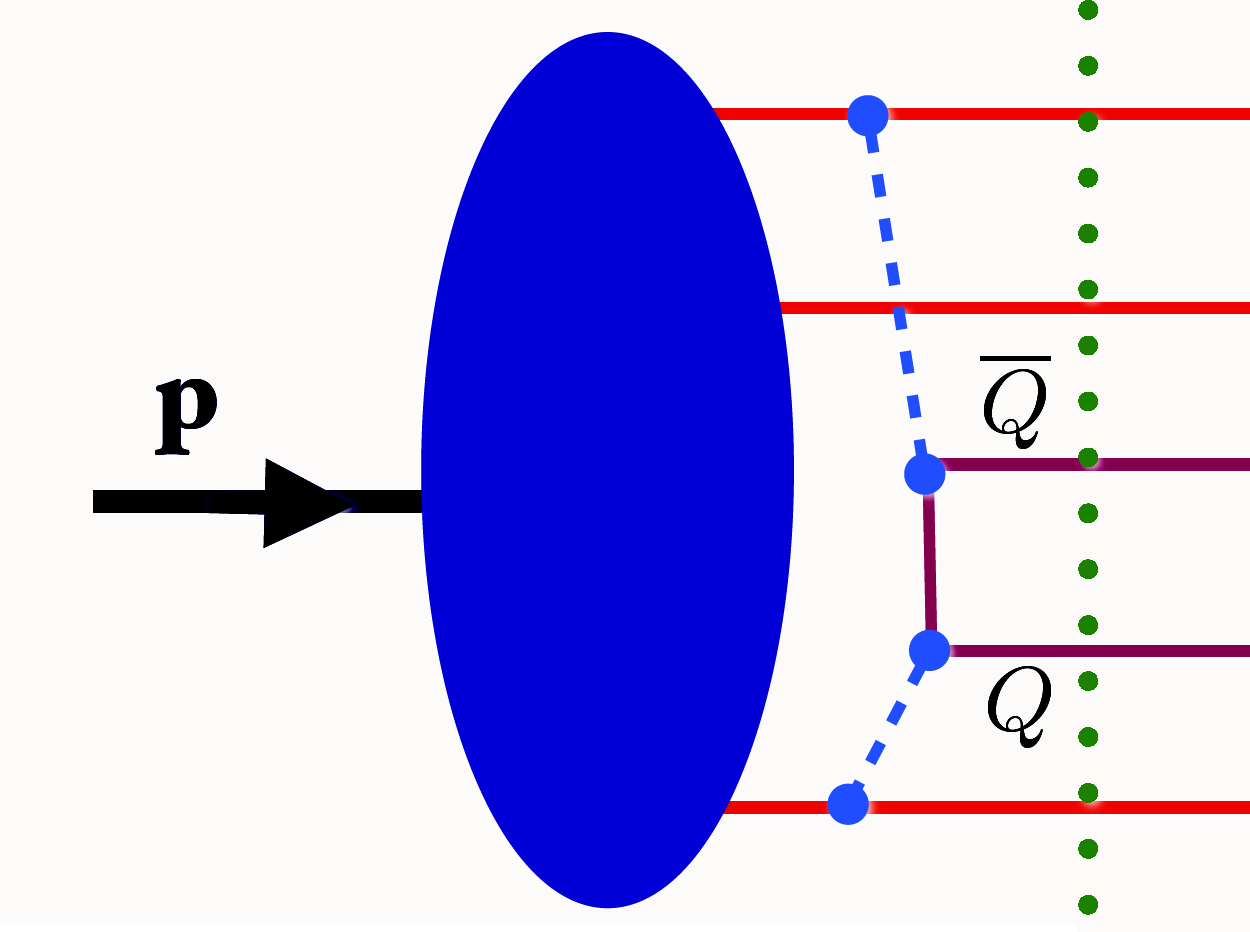}
\end{center}
\caption{Five-quark Fock state $|uudQ\overline Q\rangle$ of the proton and 
the origin of the intrinsic sea.}
\label{fig:IQ}  
\end{figure}

Production processes sensitive to the intrinsic heavy quark distributions of 
protons and nuclei are among the most interesting hadronic physics topics that 
can be investigated with AFTER@LHC. In contrast to the familiar extrinsic 
contributions which arise from gluon splitting in perturbative QCD,
the intrinsic heavy quarks have multiple connections to the valence quarks of 
the proton and thus are sensitive to its nonperturbative structure. 
For example, if the gluon-gluon scattering box diagram, 
$gg \to Q \overline Q  \to gg$ (the analog of QED light-by-light scattering), 
is inserted into the proton self-energy,  the cut of this amplitude generates 
five-quark Fock states of the proton $|uud Q\overline Q\rangle$, see  
Fig.~\ref{fig:IQ}. 

Intrinsic strange, charm, and bottom quarks are thus a fundamental property of 
the wavefunctions of hadronic bound 
states~\cite{Brodsky:1980pb,Brodsky:1984nx,Harris:1995jx,Franz:2000ee}.
While the extrinsic contributions to the heavy quark parton distribution 
functions (PDFs) are most important at low $x$ and depend logarithmically 
on the heavy quark mass $M_Q$, the intrinsic heavy quark contributions are 
dominant at high $x$ and depend on $1/M^2_Q$.  
Because the extrinsic heavy quarks are generated by gluon splitting, their
PDFs are always softer than those of the parent gluon by a factor of $(1-x)$.   
In contrast, the high $x$ intrinsic heavy quark contributions are kinematically
dominated by the regime where the $|uud Q \overline Q \rangle$ state is
minimally off shell, corresponding to equal rapidities of the 
constituent quarks.  The resulting momentum and spin distributions of the 
intrinsic $Q$ and $\overline Q$ can be distinct, {\it e.g.}, 
$s(x) \ne \overline s(x)$ since the comoving $uud Q \overline Q$ quarks
are sensitive to the global quantum numbers of the proton.

A finite intrinsic charm contribution to the nucleon has been extracted from
lattice QCD.  An analysis by the MILC collaboration~\cite{Freeman:2012ry}  yields a probability for the charm matrix element $\langle N| c\overline c |N \rangle$ in the range  of  $5 - 6$\%, consistent with a four-loop perturbative QCD calculation~\cite{Kryjevski:2003mh}.  

While the first experimental evidence of intrinsic heavy quarks came from the 
EMC measurement of the large $x$ charm structure function \cite{Aubert:1982tt},
a variety of other charm hadron and charmonium measurments are consistent
with the existence of intrinsic charm.
Open charm observables in hadroproduction include forward $\Lambda_c$ production
at the ISR \cite{Bari:1991ty}\footnote{Similarly,  the coalescence of comoving $b$, $u$ and $d$ quarks from the $|uud \bar b b>$  
intrinsic bottom Fock state 
in the proton can explain the high $x_F$ production of the $\Lambda_b(udb)$ baryon, as observed at the ISR~\cite{Bari:1991ty}.}
and asymmetries between leading and nonleading
charm ($\overline D$ mesons which share valence quarks with the projectile
and $D$ mesons which do not, respectively) measured as functions of
$x_F$ and $p_T$ in fixed-target experiments,
WA89 and WA82 at CERN; E791 and SELEX at Fermilab, see 
Refs.~\cite{Vogt:1992ki,Vogt:1995fsa,Gutierrez:1998bc} and references therein.  
Previous fixed-target $J/\psi$ measurements also give indications of 
important intrinsic charm contributions, particularly from the nuclear mass,
or $A$, dependence, as measured by NA3 at CERN as well as E772 and, later,
E866 at Fermilab, see {\it e.g.} \cite{Vogt:1991qd}.  
Indeed, the $A$ dependence, proportional to $A^\alpha$,
is quite different than the $\alpha \sim 1$ expected from extrinsic-type
production \cite{Hoyer:1990us}.  At large $x_F$, there are indications of
a $A^{2/3}$ dependence, consistent with a nuclear surface-type interaction
instead of the volume dependence of pQCD.  In addition, the NA3 collaboration
measured double $J/\psi$ production at forward $x_F$ in $\pi A$ interactions,
difficult to explain without an intrinsic charm mechanism \cite{Vogt:1995tf}.
All of these observables can be studied with higher energies and luminosities
at AFTER@LHC, making precision measurements possible for the first time.

In addition to the typical observables for intrinsic heavy quarks, these
intrinsic heavy quarks also contribute to a number of more exotic
observables and  inclusive and diffractive
Higgs production $pp \to p p H$, in which the Higgs boson carries a 
significant fraction of the projectile proton momentum 
\cite{Brodsky:2006wb,Brodsky:2007yz}.  There are also important 
implications for intrinsic charm and bottom quarks in Standard Model physics, 
as in the weak decays of the $B$-meson~\cite{Brodsky:2001yt} and a novel solution to the $J/\psi \to \rho \pi$ problem~\cite{Broadsky:2012rw}.
AFTER@LHC could also shed light on these topics.

The rest of this paper is organized as follows.
In Sec.\ \ref{sec:theory}, we give an overview of the theoretical models predicting the $x$-shape (but not the normalization)
of the intrinsic charm and bottom parton distribution functions. 
In Sec.\ \ref{sec:pdfs}, we discuss the constraints on the normalization of the intrinsic charm (IC) obtained in global analyses of PDFs.
Section \ref{sec:ib} is devoted to the intrinsic bottom (IB) content of the nucleon
for which there are currently no quantitative constraints.
In Sec. \ref{sec:observables} we review collider observables sensitive to an intrinsic charm or bottom PDF.
Finally, in Sec.\ \ref{sec:conclusions} we present our conclusions.

\section{Theoretical models}
\label{sec:theory}

The QCD wavefunction of a hadron can be represented as a
superposition of quark and gluon Fock states. For example, at fixed
light-front time, a hadron wavefunction can be
expanded as a sum over the complete basis of free quark and gluon
states: $\vert \Psi_h \rangle = \sum_m \vert m \rangle \,
\psi_{m/h}(x_i, k_{T,i})$ where the color-singlet
states, $\vert m \rangle$, represent the fluctuations in the hadron
wavefunction with the Fock components $\vert
q_1 q_2 q_3 \rangle$, $\vert q_1 q_2 q_3 g \rangle$, $\vert
q_1 q_2 q_3 c \overline c \rangle$, {\it etc}. The boost-invariant
light-front wavefunctions, $\psi_{m/h}(x_i, k_{T,i})$
are functions of the
relative momentum coordinates $x_i = k_i^+/P^+$ and $k_{T,i}$
where $k_i$ denotes the parton momenta and $P$ the hadron momentum.
Momentum conservation demands $\sum_{i=1}^n x_i = 1$ and
$\sum_{i=1}^n \vec{k}_{T,i}=0$ where $n$ is the number of partons
in state $\vert m \rangle$.
For example, 
as predicted by Brodsky and collaborators, in the BHPS model 
intrinsic charm
fluctuations \cite{Brodsky:1980pb,Brodsky:1981se} can be liberated by a soft interaction which breaks
the coherence of the Fock state \cite{Brodsky:1991dj}
provided the system is probed during
the characteristic time that such fluctuations exist.

Microscopically, the intrinsic heavy quark Fock component in the
proton wavefunction, $|u u d c \overline c \rangle$, is
generated by virtual interactions such as $g g \rightarrow Q
\overline Q$ where the gluons couple to two or more
valence quarks. The probability for $c \overline c$ fluctuations to
exist in a hadron is higher twist since it scales as $1/m_c^2$
relative to the extrinsic, EC,
leading-twist production by photon-gluon fusion \cite{Vogt:1995tf}.

The dominant Fock state configurations are not far off
shell and thus have minimal invariant mass, $M^2 = \sum_i^n \widehat{m}_i^2/
x_i$ where  $\widehat{m}_i^2 = m_i^2 + \langle
\vec k_{T, i}^2 \rangle$ is the square of the average transverse mass
of parton $i$.
The general form of the Fock state wavefunction for a hadron with mass $m_h$ 
appropriate to any frame at fixed light-front time is \be
\Psi(x_i, \vec k_{\perp i}) = \frac{\Gamma(x_i, \vec k_{\perp i}) }{m_h^2 -
M^2 } \, \,  \ee where $\Gamma$ is a
vertex function, expected to be a slowly-varying,
decreasing function of $m_h^2 - M^2$.
The particle distributions are then
controlled by the light-front energy denominator and  phase space.
This form for the higher Fock components is applicable to an
arbitrary number of light and heavy partons.
Intrinsic $c \overline c$ Fock components with minimum invariant
mass correspond to configurations with equal rapidity constituents.
Thus, unlike extrinsic heavy quarks generated from a single parton, intrinsic
heavy quarks carry a larger fraction of the parent momentum
than the light quarks in the state \cite{Brodsky:1980pb,Brodsky:1981se}.

The parton distributions
reflect the underlying shape of the Fock state wavefunction.
Assuming it is sufficient to
use $\langle k_T^2 \rangle$ for the transverse momentum,
the probability distribution as a function of $x$ in 
a general $n$--particle intrinsic
$c \overline c$ Fock state is
\be
\label{icprobtot}
\frac{dP_{\rm IC}}{dx_i \cdots dx_n} =  N_n 
%\alpha_s^4(M_{c \overline c})
\ \frac{\delta(1-\sum_{i=1}^n x_i)}{(m_h^2 - \sum_{i=1}^n
(\widehat{m}_i^2/x_i)
)^2} \, \, ,
\ee
where $N_n$ normalizes the $n$-particle Fock state probability.  

At LO in the
heavy quark limit, $\widehat{m}_c$, $\widehat{m}_{\overline c} \gg m_h$,
$\widehat{m}_q$, 
\begin{equation}
\frac{dP_{\rm IC}}{dx_i \cdots dx_n}  =  N_n 
%\alpha_s^4(M_{c \overline c}) 
\frac{x_c^2 x_{\overline c}^2}{(x_c + x_{\overline c})^2}
\ \delta\Big(1-\sum_{i=1}^n x_i\Big) \, ,
\label{massless1}  
\end{equation}
leading to
\begin{eqnarray}
F_{2 \, c}^{\rm IC \, LO}(x) & = & 
\frac{8}{9} xc(x) 
\nonumber\\
&=& \frac{8}{9}x \int dx_1 \cdots
dx_{\overline c} \frac{dP_{\rm IC}}{dx_i \cdots dx_{\overline c}
dx_c} \, \, .
\label{massless2}  
\end{eqnarray}

There are many applications of intrinsic charm in
charm hadron production.  See, e.g., 
Refs.~\cite{Vogt:1995tf,Vogt:1991qd,Vogt:1992ki,Vogt:1995fsa,Gutierrez:1998bc}
for more details.

Paiva {\it et al.} have also calculated an intrinsic charm component of the
nucleon sea within the context of the meson cloud model \cite{Paiva:1996dd}.  They
assumed that the nucleon can fluctuate into $\overline D \Lambda_c$.  The
$\overline c$ distribution in the nucleon is then
\begin{equation}
x {\overline c}_N (x) = \int_{x}^{1}
dy\, f_{\overline D} \left(y\right)\, \frac{x}{y} \, {\overline c}_{\overline D} 
\left(\frac{x}{y}\right)\; .  
\label{cn}
\end{equation}                       
where
\begin{equation}
f_{\overline D} (y) = \frac{g^2_{ \overline D N\Lambda_c}}{16 \pi^2} \, y \, 
\int_{-\infty}^{t_{\rm max}}dt \, \frac{[-t+(m_{\Lambda_c}-m_N)^2]}{[t-
m_{\overline D}^2]^2}\,
F^2 (t)\; ,
\label{fdbar}
\end{equation}
with $F(t)$ a form factor at the $DN\Lambda$ vertex
and $t_{\rm max} = m^2_N y- m^2_{\Lambda_c} y/(1-y)$.  In this case they 
chose a monopole form factor with $\Lambda_m = 1.2$ GeV.  The coupling 
constant was assumed to be $g_{\overline D N \Lambda_c} = -3.795$.  From heavy
quark effective theories \cite{Neubert:1993mb}, the $\overline c$ distribution in the 
$\overline D$ is expected to be hard because in the bound state, the $\overline
c$ exchanges momenta much less than $m_c$.  They make the extreme assumption
that the entire $\overline D$ momentum is carried by the charm quark,
$\overline c_{\overline D} = x \delta(x-y)$.

Next, Steffens {\it et al.}\ investigated all 
the charm structure function
data with two variants of intrinsic charm \cite{Steffens:1999hx}.  The first was that of 
Eq.~(\ref{massless2}), called IC1 in their paper, while the second was a meson
cloud model, IC2.  In the second approach, the $\overline c$ distribution is
obtained from the light-front distribution of $\overline D^0$ mesons in the
nucleon,
\begin{eqnarray}
\overline c^{\rm IC2}(x) & \approx & f_{\overline D}(x) 
 =  \frac{1}{16\pi^2}
\int_0^\infty dk_\perp^2 \frac{g^2(x,k_\perp^2)}{x(1-x)(s_{\overline D 
\Lambda_c}-m_N^2)^2} 
\nonumber\\
&&\times \frac{k_\perp^2 + (m_{\Lambda_c} - (1-x)m_N)^2}{1-x} \, \, .
\label{cbaric2} 
\end{eqnarray}
A hard charm momentum distribution was assumed in the $\overline D$, similar to
that of Ref.~\cite{Paiva:1996dd}.
The vertex function $g^2(x,k_\perp^2)$ is parameterized as $g^2 =
 g_0^2(\Lambda^2 + m_N^2)/(\Lambda^2 + s_{\overline D \Lambda_c})$ where
$ s_{\overline D \Lambda_c}$ is the square of the center of mass energy of
the $\overline D \Lambda_c$ system and $g_0^2$ the coupling constant at
$ s_{\overline D \Lambda_c} = m_N^2$.  For an intrinsic charm probability of
1\%, $\Lambda \approx 2.2$ GeV.  The charm distribution is then
\be
c^{\rm IC2}(x) \approx \frac{3}{2} f_{\Lambda_c} \left(\frac{3x}{2} \right)
\label{cic2}
\ee
where the charm distribution in the $\Lambda_c$ is assumed to be $c_{\Lambda_c}
\sim \delta(x - 2/3)$ and $f_{\Lambda_c}(x) = f_{\overline D}(1-x)$.

Pumplin \cite{Pumplin:2005yf} considered a model where a point scalar particle of
mass $m_0$ couples with strength $g$ to $N$ scalar particles with mass
$m_1$, $m_2$, $\cdots$, $m_N$.  The probability density is then
\begin{eqnarray}
dP &=& \frac{g^2}{(16\pi^2)^{N-1}(N-2)!} \prod_{j=1}^N dx_j \delta \bigg(1 - 
\sum_{j=1}^N x_j \bigg) \times
\nonumber\\
&& \int_{s_0}^\infty ds \frac{(s - s_0)^{N-2}}{(s - m_0^2)^2}
|F(s)|^2\, ,
\label{Pumplin_eq}
\end{eqnarray}
where $s_0 = \sum_{j=1}^N (m_j^2/x_j)$.  The form factor $F(s)$ suppresses
higher mass state contributions.  If the quark transverse momenta are neglected,
with $m_c$ much greater than all other mass scales, and $F(s) = 1$, then
the BHPS model is recovered.  Two types of form factors were studied,
an exponential $|F(s)|^2 = \exp[-(s-m_0^2)/\Lambda^2]$, and a power law,
$|F(s)|^2 = 1/(s + \Lambda^2)^n)$ where the cutoff $\Lambda$ is varied between 2 and 10
GeV.

Hobbs {\it et al.} employed a meson cloud type approach but specified the
spin and parity of all lowest mass charm meson-baryon combinations from the
5-particle $|uudc \overline c \rangle$ Fock states of the proton \cite{Hobbs:2013bia}.  
They pointed out that treating quarks as 
scalar point-like particles, as in {\it e.g.} Ref.~\cite{Pumplin:2005yf}, does not
conserve spin and parity.  They calculated the appropriate meson-baryon
splitting functions for the meson-baryon combinations and found that the
production of charm mesons would be almost entirely through $D^*$ mesons.
To study the phenomenological distributions of charm mesons and baryons in
this approach, they studied exponential and confining vertex functions,
$\propto \exp[-(s - m_D^2)/\Lambda^2]$ and 
$(s-m_D^2)\exp[-(s - m_D^2)/\Lambda^2]$ respectively.  They used these results
to compare to the $\Lambda_c$ distribution from the  ISR \cite{Chauvat:1987kb} and the 
$\Lambda_c/\overline \Lambda_c$ asymmetry from SELEX \cite{Garcia:2001xj}.  See 
Ref.~\cite{Hobbs:2013bia} for details.

\section{Global analyses of PDFs with intrinsic charm}
\label{sec:pdfs}

In the standard approach employed by almost all global analyses of PDFs, the heavy quark distributions 
are generated {\em radiatively}, according to DGLAP evolution equations~\cite{Altarelli:1977zs,Gribov:1972ri,Dokshitzer:1977sg}, 
starting with a perturbatively calculable boundary condition 
\cite{Collins:1986mp,Buza:1996wv} 
at a scale of the order of the heavy quark mass.
In other words, there are no free fit parameters associated to the heavy quark distribution and
it is entirely related to the gluon distribution function at the scale of the boundary condition. 
As a consequence, also the PDF uncertainties for the heavy quark and the gluon PDFs are strongly 
correlated as has been discussed in the context of inclusive Higgs production at the Tevatron and
the LHC \cite{Belyaev:2005nu}.
However, a purely perturbative treatment might not be adequate,
in particular for the charm quark with a mass $m_c \simeq 1.3$ GeV
which is not much bigger than typical hadronic scales
but also for the bottom quark with a mass $m_b \simeq 4.5$ GeV.
Indeed, as discussed above, light-front models predict a nonperturbative ('intrinsic') heavy quark
component in the proton wave-function~\cite{Brodsky:1980pb,Brodsky:1981se}.
Motivated by the theoretical predictions of the BHPS light-front model, analyses
of the charm distribution in the proton
going beyond the common assumption of purely radiatively generated charm
date back almost as far as the BHPS predictions themselves.
For definiteness, in the following we refer to the radiatively generated charm by $c_0(x,Q)$
and to the intrinsic charm by $c_1(x,Q)$. The full charm parton distribution is then given
by the sum $c(x,Q)=c_0(x,Q)+c_1(x,Q)$. Strictly speaking, this decomposition is defined at the initial scale 
$Q_0 \simeq m_c$ of the DGLAP evolution but holds to a good approximation at any scale
since the intrinsic component $c_1$ is governed (to a very good approximation) by a
standalone non-singlet evolution equation \cite{Lyonnet:2015dca}.
A similar decomposition is understood for the bottom quark which will be discussed in Sec.\ \ref{sec:ib}.

The BHPS model of the $|uud c \overline c \rangle$ Fock state predicts a
simple form for $F_{2 \, c}(x)$,
\begin{eqnarray}
\label{eq:IC}
F_{2\, c}^{\rm IC}(x) &=& \left(\frac{8}{9} x\right) \frac{1}{2} N_5 x^2 \times
\\
&&
\left[\frac{1}{3}(1-x)(1 +
10x + x^2)+ 2x(1+x)\ln x\right] \, \, .
\nonumber
\end{eqnarray}
If there is a 1\% intrinsic charm contribution to the proton PDF, $N_5 = 36$.

Hoffman and Moore incorporated mass effects and introduced next-to-leading
order corrections as well as scale evolution \cite{Hoffmann:1983ah}.  They
compared their result to the EMC $F_{2 \, c}$ data from muon scattering on iron
at high $x$ and $Q^2$ with
the intrinsic charm contribution added to the leading order calculation of
$F_{2 \, c}$ by photon-gluon fusion.  

A complete next-to-leading order analysis
of both the `extrinsic' radiatively-generated charm component and the 
intrinsic component was later carried out by Harris {\it et al.} 
\cite{Harris:1995jx}.
The EMC data with $\overline{\nu} = \overline{Q^2} / 2m_p 
\overline{x} = 53,\, 95, \,$ and
168 GeV were fit by a sum of the extrinsic and intrinsic components 
\cite{Harris:1995jx}.  The normalization
of the two components were left as free parameters,
\begin{eqnarray} 
\label{f2ccombo}
F_{2 \,c}(x,\mu^2,m_c^2) &=& \epsilon  F_{2 \, c}^{\gamma p}(x,\mu^2,m_c^2)
                   + \delta  F_{2 \, c}^{\rm IC}(x,\mu^2,m_c^2)\, ,
\nonumber\\
\end{eqnarray}
with the scale $\mu = \sqrt{m_{c \overline c}^2 + Q^2}$.  
The parameter $\epsilon$, typically larger than unity, was considered to be 
an estimate of the NNLO contribution to the extrinsic contribution.
Since a 1\% normalization of the IC component was assumed in 
Eq.~(\ref{f2ccombo}), the fitted value of $\delta$ is the fraction of 
this normalization.  Given the quality of the data, no
statement could be made about the intrinsic charm content of the proton when
$\bar{\nu}=53$ and $95 \: {\rm GeV}$. However,
with $\bar{\nu}=168 \: {\rm GeV}$
an intrinsic charm contribution of $(0.86\pm0.60)\%$ was indicated.
These results were consistent with those of the original analysis by Hoffman
and Moore \cite{Hoffmann:1983ah}.

The BHPS light-front model assumes that $c_1(x) = \overline c_1(x)$.  Meson cloud
models, introduced later, treat the 5-particle Fock state as a combination
of (predominantly) $\overline D^0 \Lambda_c^+$.  In this case, of course,
$c_1(x) \neq \overline c_1(x)$ with the $\overline c$ quark in the $\overline D^0$
carrying more momentum than the $c$ quark in the charm baryon.  An analysis
by Steffens {\it et al.} in the context of the meson cloud model and using
a hybrid scheme to interpolate between massless evolution at high $Q^2$
and `extrinsic' production at low $Q^2$ found a limit of $\sim 0.4$\%
\cite{Steffens:1999hx}.

Regardless of
whether or not the models predict $\overline c_1(x) - c_1(x) > 0$, intrinsic
charm should provide the dominant contribution to the charm density in the
proton at large $x$ \cite{Pumplin:2005yf}. 

For some time, no other analyses of the charm structure function were made.
The EMC data remain the only measurement of the charm structure function in
the relevant $(x,Q^2)$ regime and are the only DIS data cited as evidence
for intrinsic charm.  The HERA data on $F_{2 \, c}$ were at too low $x$ to
address the issue.  

The first global analyses of the proton PDFs with an
intrinsic charm contribution included were performed by members of the CTEQ 
collaboration \cite{Pumplin:2007wg,Nadolsky:2008zw}.
In addition to the BHPS and meson cloud approaches, they also allowed for a
`sea-like' contribution with the same shape as the radiatively-generated charm
distribution.  
They characterized the magnitude of the intrinsic charm component ($c_1(x,Q^2)$)
by the first moment of the charm distribution at the input scale $Q_0=m_c=1.3$ GeV:\footnote{Note that at $Q_0=m_c$ the radiatively
generated charm component ($c_0(x,Q^2)$) vanishes at NLO in the ${\overline{\rm MS}}$ scheme 
so that $c(x,Q_0^2)=c_1(x,Q_0^2)$.}
\begin{equation}
c_1(N=1,Q_0^2) = \int_0^1 dx\ c_1(x,Q_0^2) = 0.01\, ,
\label{eq:norm}
\end{equation}
which translates into a momentum fraction
\begin{equation}
\langle x \rangle_{c_1 + \overline c_1} = \int_0^1 dx \, x[c_1(x,Q_0^2) + \overline c_1(x,Q_0^2)] = 0.0057 \,\, . 
\label{Eq:avex}
\end{equation}
They found that the global analyses of hard-scattering data provided no evidence
for or against the existence of intrinsic charm up to $\langle x 
\rangle_{c_1 + \overline c_1} = 0.0057$, {\it i.e.} the quality of the fit is
insensitive to $\langle x \rangle_{c_1 + \overline c_1}$ in this interval.  They also
found that the allowed range was greatest for the sea-like IC, expected since
this shape is rather easily interchangeable with other sea quark components
while the other, harder, charm distributions are not \cite{Pumplin:2007wg}.
In addition, they concluded that the enhancement due to IC relative to analyses without
it persisted up to scales of $\sim 100$ GeV and could have an influence on
charm-initiated processes at the LHC, as is discussed later.  The CTEQ6.6C
proton PDFs were generated as a result of this analysis \cite{Nadolsky:2008zw}.

There are two recent updates to the global analyses, reaching different
conclusions about the importance of intrinsic charm.  The first, by Dulat
{\it et al.} \cite{Dulat:2013hea}, follows the previous work in the context
of the CTEQ collaboration \cite{Pumplin:2007wg,Nadolsky:2008zw}.  The second,
by Jimenez-Delgado {\it et al.} \cite{Jimenez-Delgado:2014zga}, 
included more lower energy data
than the previous global analyses.

The result of Dulat {\it et al.} \cite{Dulat:2013hea}
was based on the CT10 NNLO parton densities.  Here the strong coupling,
$\alpha_S(Q^2)$, the evolution equations and the matrix elements are calculated
at NNLO.  Only the inclusive jet data still required NLO expressions.
Their analysis included DIS data from BCDMS, NMC, CDHSW, and CCFR; SIDIS data
from NuTeV and CCFR; the combined DIS and $F_{2\,c}$ data from HERA; Drell-Yan
production; the $W$ charge asymmetry and $Z^0$ rapidity from CDF and D0; and
the inclusive jet measurements from CDF and D0, see Ref.~\cite{Dulat:2013hea}
for a complete list.  

\begin{figure}[thp!]
\centering
\includegraphics[width=0.45\textwidth]{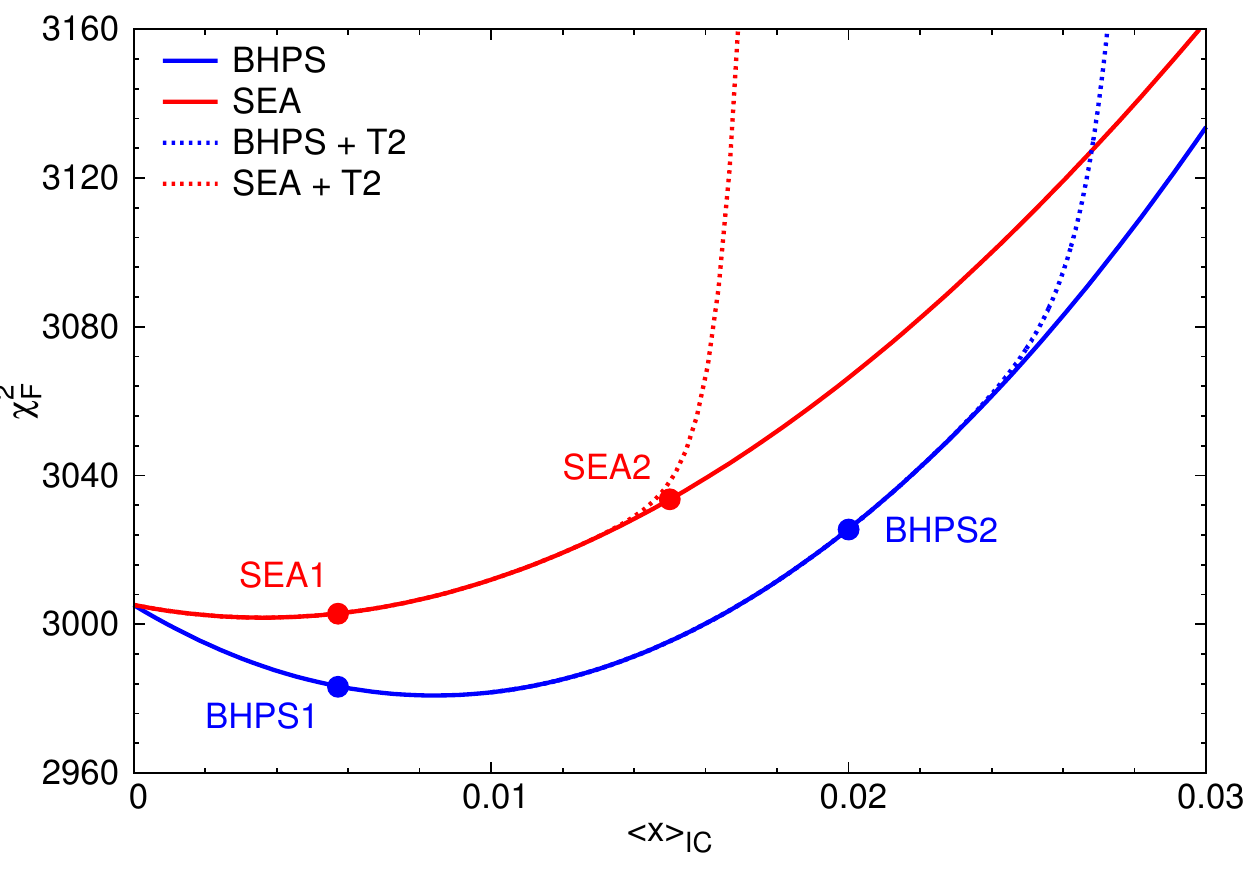}
\caption{ (Color online) The global chi-square function versus charm momentum fraction
$\langle x \rangle_{\rm IC}$. The two curves are determined from fits with
many values of $\langle x \rangle_{\rm IC}$. Two exemplary fits for each
IC model are shown as dots. Blue dots denotes the BHPS model; the dots have
$\langle x \rangle_{\rm IC}=0.57\%$ and 2\%, which are denoted as BHPS1 and
BHPS2. Red denotes SEA model; the dots have $\langle x \rangle_{\rm IC}=0.57\%$
and 1.5\%, which are denoted SEA1 and SEA2.
Additionally the dotted lines show global chi-square function with additional
penalty, $T_2(i)$, used to set the upper limits on the allowed IC component.
\\(Figure taken from \cite{Dulat:2013hea})}
\label{fig:Dulat}
\end{figure}

Two models of IC were considered: the BHPS light-front
model and the sea-like IC introduced in Ref.~\cite{Pumplin:2007wg}.
They found a broader possible probability range for IC in this analysis,
$\langle x \rangle_{\rm IC} = \langle x \rangle_{c_1 + \overline c_1}(Q_0^2) \lesssim 0.025$ 
for BHPS and $\langle x \rangle_{\rm IC} \lesssim 0.015$ for the sea-like IC,
summarized in Fig.~\ref{fig:Dulat}.
This finding differs from the previous work which found a larger upper limit
on IC for the sea-like model. They believe that the difference is caused
by the improved treatment of the charm quark
mass in the later study \cite{Dulat:2013hea}.
%--------------
%--------------

In addition to the global fit, they also tested the
sensitivity of their result to individual experiments by introducing a penalty
factor, $T_2(i)$, for each experiment $i$.  This penalty factor is designed to
increase more rapidly than the $\chi^2_i$ for that experiment when $\chi^2_i$
goes beyond the 90\% confidence level.  The penalty factor employs an
equivalent Gaussian variable $S_n$ which measures the goodness of fit for
each individual data set.  Values of $S_n \leq |1|$ are considered good fits,
$S_n > 3$ is considered to be a poor fit, and values of $S_n < -3$ are better
fits than expected from usual statistical analyses.  Using the $S_n$ dependence
on $\langle x \rangle_{\rm IC}$, they determined which of the data sets used in
the global analyses are most sensitive to intrinsic charm.  The upper limit
on the BHPS value of $\langle x \rangle_{\rm IC}$ comes from the CCFR structure
function data while the HERA combined charm data sets the upper limit on IC
from the sea-like model \cite{Dulat:2013hea}.  

They also studied the sensitivity of their sea-like result to the charm quark
mass and found that, if the charm quark mass was raised from 1.3 GeV, as in
the CT10 fits, to 1.67 GeV, then the minimum $\chi^2$ for the global analyses
would support $\langle x \rangle_{\rm IC} = 0.01$ rather than 0 although the
global $\chi^2$ is worse for the larger charm mass \cite{Dulat:2013hea}.
Finally, they showed how $W$ and $Z$ production at the LHC might be affected
by a nonzero IC contribution.

In the most recent study, Jimenez-Delgado {\it et al.} 
\cite{Jimenez-Delgado:2014zga} included the full range of high energy scattering
data by using looser kinematic cuts $Q^2 \geq 1$ GeV$^2$ and 
$W^2 \geq 3.5$ GeV$^2$. In particular, they included the lower energy SLAC fixed-target data
which did not pass the more stringent standard DIS cuts on the $(Q^2,W^2)$ plane
applied in the previous work \cite{Pumplin:2007wg,Nadolsky:2008zw,Dulat:2013hea}.
The EMC $F_{2\,c}$ data, cited as the strongest
evidence for intrinsic charm in DIS, are used as a consistency check. 
The low energy, high-$x$, fixed target data lie precisely in the region
where IC is expected to be most important.
Thus including these data could enhance the
sensitivity of the global fit to IC.  Note, however, that some of these 
newly-added data are on heavier targets than the deuteron and thus target
mass corrections, nuclear corrections for $A>2$, and higher-twist effects
need to be taken into account \cite{Jimenez-Delgado:2014zga}.

They followed the framework of the JR14 \cite{Jimenez-Delgado:2014twa} global fit which decomposed
$F_2$ into light and heavy components.  The charm component is itself separated
into the 'extrinsic' and intrinsic charm components.  The fixed-flavor number
scheme is used to compute the extrinsic contribution.  In this scheme, the
charm quark mass enters the PDF evolution only indirectly through the running
of $\alpha_s$ \cite{Jimenez-Delgado:2014zga}.  They employed a charm quark mass
of 1.3 GeV, as did Dulat {\it et al.} \cite{Dulat:2013hea}.  They used all
three intrinsic charm models previously considered: BHPS, the meson-cloud 
model (this time including pseudoscalar and vector mesons as well as spin $1/2$
and spin $3/2$ charm baryons -- the CTEQ analyses only included the scalar
$\overline D \Lambda_c$ fluctuation), and the sea-like component 
\cite{Jimenez-Delgado:2014zga}.  The IC contribution was evolved up to NLO.

\begin{figure}[thp!]
\centering
\includegraphics[width=0.45\textwidth]{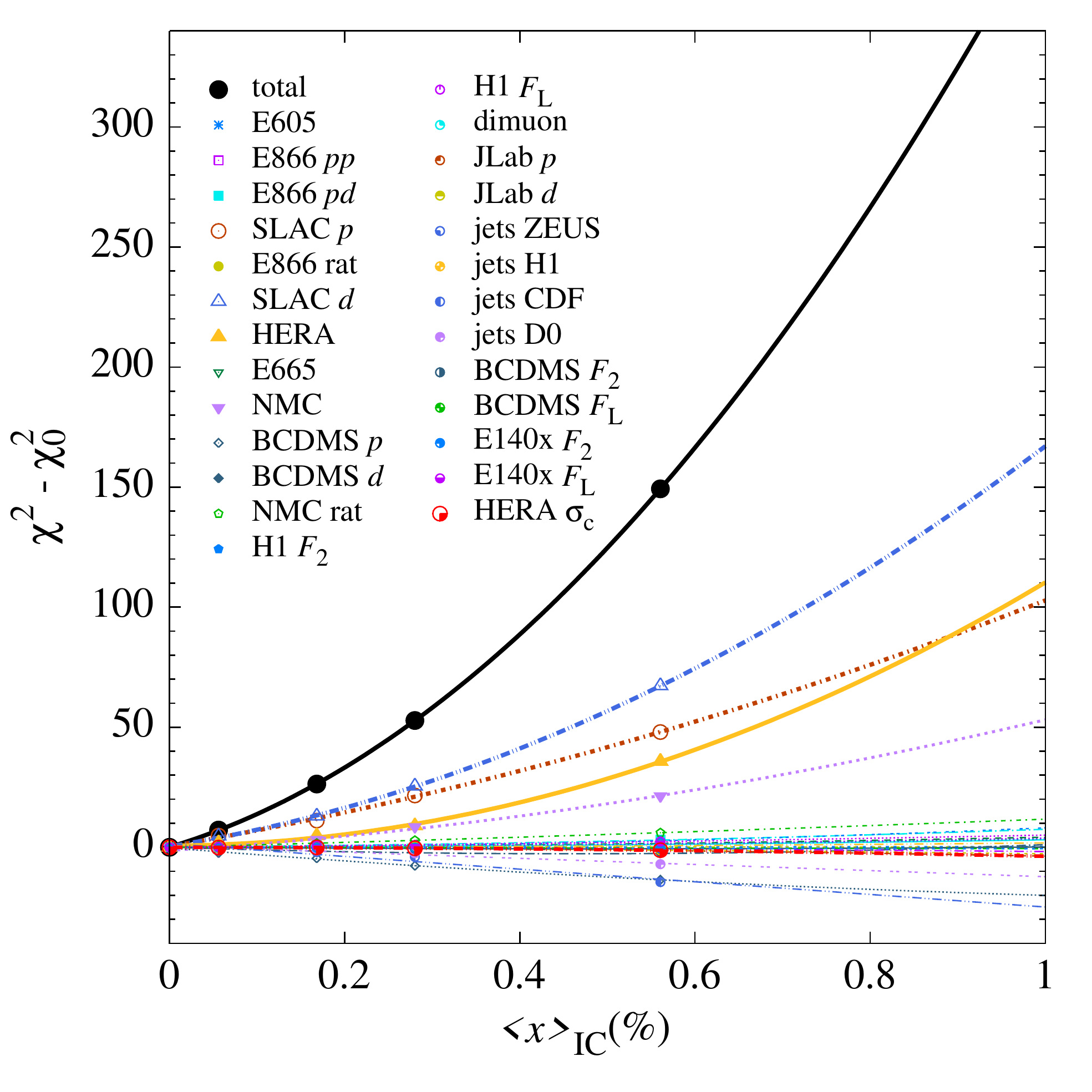}
\caption{ (Color online) Contributions to the total $\chi^2$ (black circles), relative
to the value $\chi_0^2$ for no IC, of various data sets as a function of the
momentum fraction $\langle x \rangle_{\rm IC}$.\\
(Figure taken from \cite{Jimenez-Delgado:2014zga})}
\label{fig:Delgado}
\end{figure}

They found that the total $\chi^2$ is minimized for 
$\langle x \rangle_{\rm IC} = 0$ with $\langle x \rangle_{\rm IC} < 0.1$\% at
the $5\sigma$ level.  When a hadron suppression factor to suppress charm
contributions near threshold is applied, they find a minimum $\chi^2$ at
$\langle x \rangle_{\rm IC} = (0.15 \pm 0.09)$\% for the full data set.
The SLAC $F_2$ (large $x$), NMC cross sections (medium $x$)
and HERA $F_{2\, c}$ (low $x$) display the greatest sensitivity to IC,
see Fig.~\ref{fig:Delgado} for details.
However, fits without the SLAC data still give a low
IC contribution \cite{Jimenez-Delgado:2014zga}.  The difference between their
results and previous results is in part due to the very different tolerance
criteria, $\Delta \chi^2 = 1$ for their fit and $\Delta \chi^2 = 100$ for
Dulat {\it et al.} \cite{Dulat:2013hea}.  Increasing the tolerance to
$\Delta \chi^2 = 100$ would also accommodate $\langle x \rangle_{\rm IC} = 1$\%
at the $1\sigma$ level \cite{Jimenez-Delgado:2014zga}.\footnote{For a critical discussion of the 
analysis in \protect\cite{Jimenez-Delgado:2014twa} and in particular of the tolerance criterion
$\Delta \chi^2 = 1$ see Ref.~\protect\cite{Brodsky:2015uwa}.}

%--------------
%--------------

When checked against the EMC $F_{2\, c}$ data, a clear preference for IC is
found, as expected, for the highest-$x$ data.  However, these data are typically
not included in global analyses due to their greater tension with other global
data sets.  

Given that the two most recent analyses set significantly different limits on
IC, it is important to collect further large-$x$ data, particularly on
$F_{2\, c}$ to try and place greater confidence on the limit of IC in the 
nucleon.  This would be an important measurement at the future electron-ion
collider.

\section{Predictions for intrinsic bottom}
\label{sec:ib}
In contrast to the case of intrinsic charm, there is currently no global analysis available that 
investigates the possibility of an intrinsic bottom (IB) content of the nucleon. 
The main reason for this is the lack of experimental data that could constrain it. 
The BHPS light-front model \cite{Brodsky:1980pb} predicts the existence of 
IB with an $x$-shape very similar to the one of IC given in Eq.~\eqref{eq:IC}
but with a normalization which is parametrically suppressed by the ratio
$m_c^2/m_b^2$. 
This fact, together with the observation that the IB PDF is governed (to an excellent approximation)
by an independent non-singlet evolution equation \cite{Lyonnet:2015dca}, 
can be used to investigate IB in a flexible way without the need of a dedicated global analysis.
Such a study has been done in Ref.~\cite{Lyonnet:2015dca} where a set of decoupled IB (and IC) PDFs has been 
provided and used together with the CTEQ6.6 PDFs~\cite{Nadolsky:2008zw} to estimate the impact of
the IB on new physics searches at the LHC. The advantage of this approach is that
the provided IB (IC) PDF can be used with any standard set of PDFs and the normalization
of the intrinsic component can be freely adjusted. This is especially useful for studies
of possible IB effects, as in that case, there are no experimental limits on what amount
of IB is allowed.

\begin{figure}[thp!]
\centering
\includegraphics[width=0.45\textwidth]{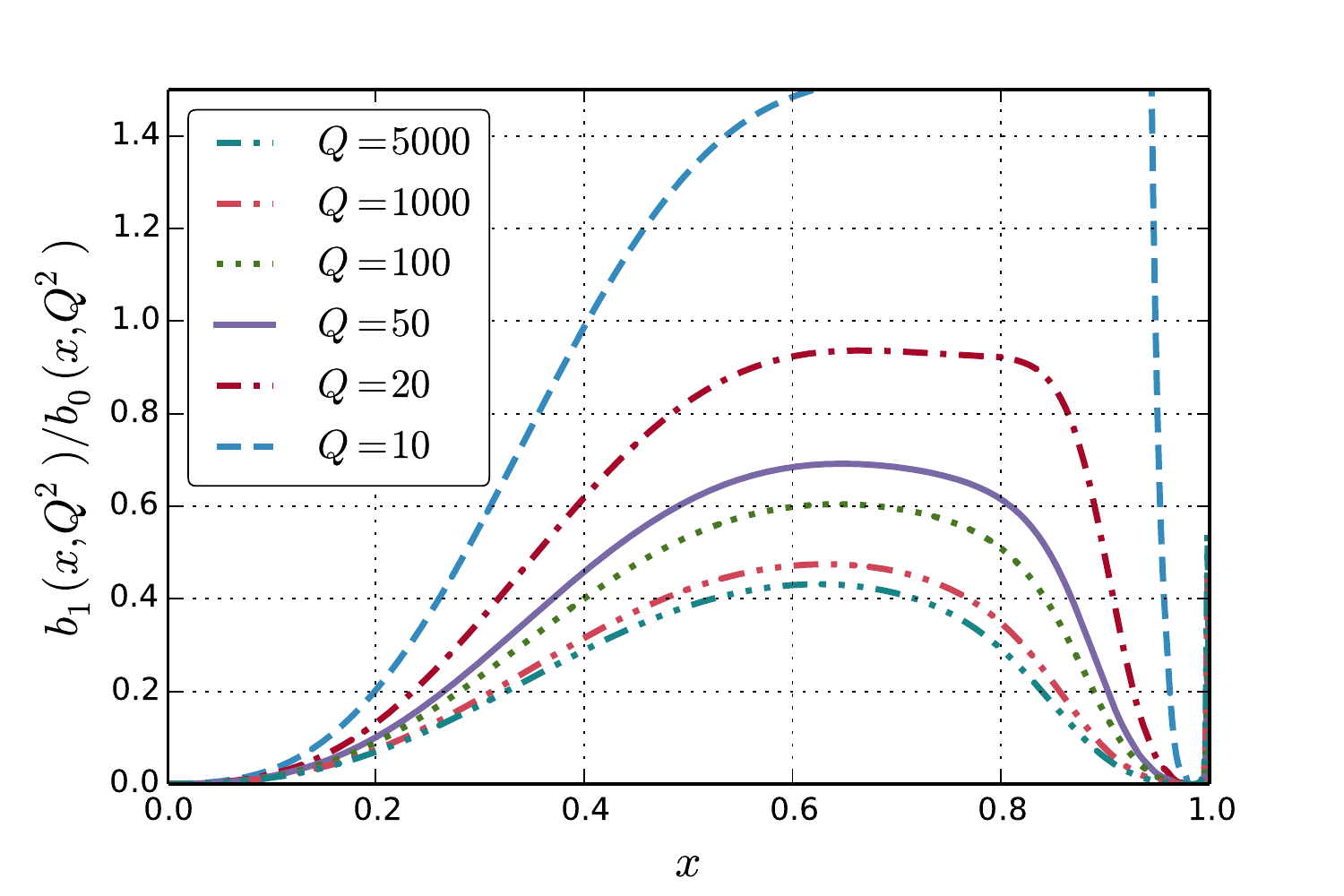}
\caption{ (Color online) Ratio of intrinsic ($b_1$) and dynamically generated ($b_0$) bottom PDFs
for various $Q$ scales. The perturbative bottom PDF from CTEQ6.6c0 \cite{Nadolsky:2008zw}
is used, the normalization of the IB is taken to be such that $\int_0^1 dx b_1(x,m_c^2) = 0.01 \times m_c^2/m_b^2$.
\\(Figure taken from \cite{Lyonnet:2015dca})
}
\label{fig:b1-b0}
\end{figure}

In the following we show some of the results found in Ref.~\cite{Lyonnet:2015dca}.
In this work, the boundary condition for the IB distribution was modeled using the IC distributions
in the CTEQ analyses \cite{Pumplin:2007wg,Nadolsky:2008zw} scaled down by
the mass factor $m_c^2/m_b^2$. 
The result of such an intrinsic bottom distribution $b_1(x,Q^2)$, with normalization 
$\int_0^1 dx b_1(x,m_c^2) = 0.01\times m_c^2/m_b^2$, 
is shown in Fig.~\ref{fig:b1-b0}, where the ratio of the intrinsic ($b_1$) and the radiatively generated ($b_0$)
component of the bottom PDF is plotted.
%--------------
%--------------
As always in the light-front models the intrinsic component is mostly present at
large $x$ values. We can see that for low scales $Q\sim10$ GeV the modification
of the bottom PDF, $\kappa_b=1+b_1/b_0$, can reach $\kappa_b=2.5$. However,
it decreases rapidly with the rising scale.
Since $b_1$ evolves independently of the
other PDFs the change in the normalization of the IB component in Fig.~\ref{fig:b1-b0}
can be done by simply rescaling the curves in the figure. If we allowed for a
$0.035 \times m_c^2/m_b^2$ normalization of the IB the modification of the bottom PDF
would be given by $\kappa_b=1+b_1/b_0\times3.5$, which for high $x$ and $Q\sim10$ GeV
would result in an enhancement of the bottom PDF by a factor  $\sim6.25$. 
However, at a scale of around 100 GeV and $x$ below 0.2-0.3, even with the
higher IB normalization, the effect is becoming negligible.

\begin{figure}[thp!]
	\centering
	\includegraphics[width=0.45\textwidth]{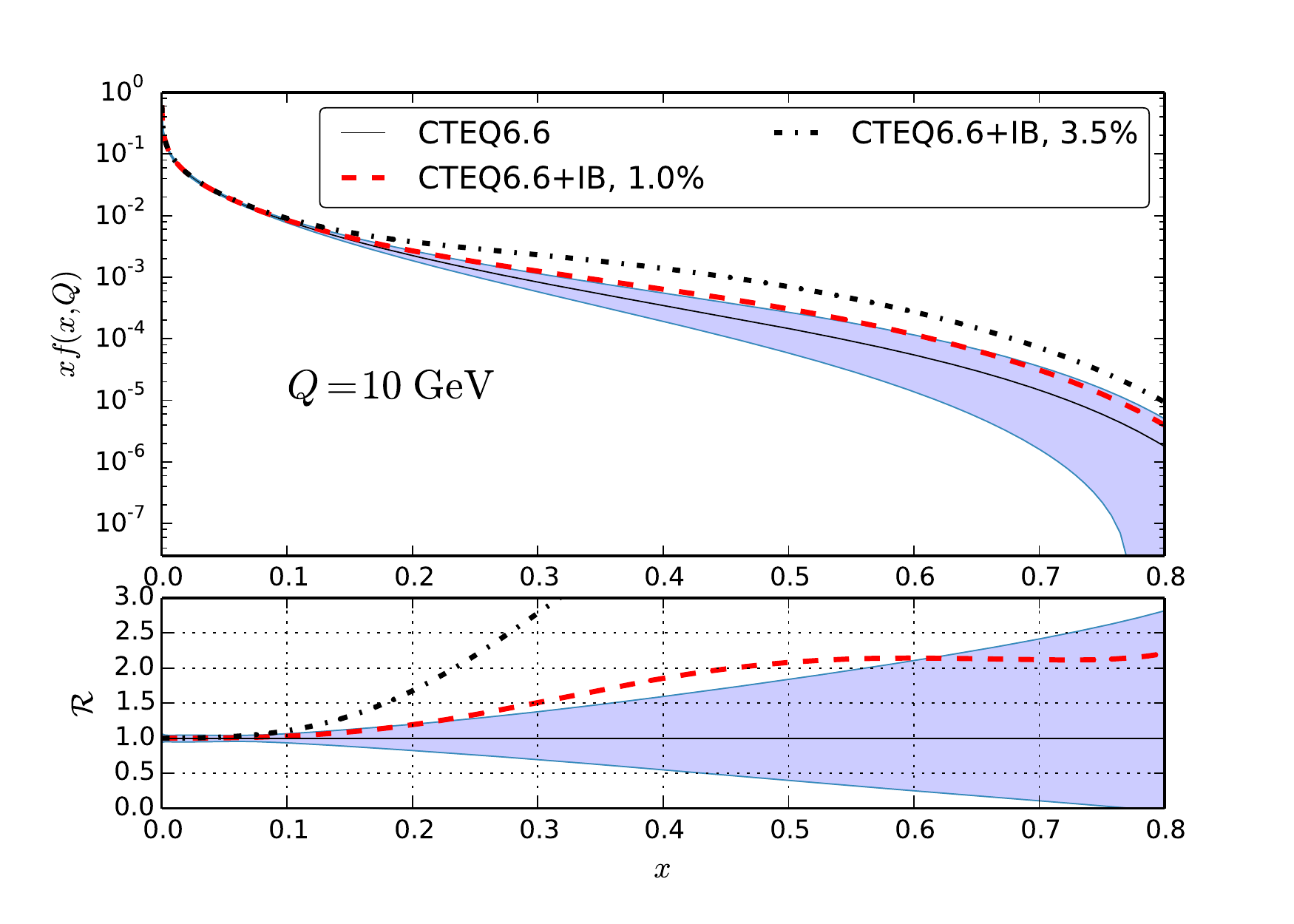}
	\caption{ (Color online) CTEQ6.6 + $b_1$ for different normalizations of the intrinsic bottom-quark PDF at the scale $Q=10$ GeV, compared to the asymmetric PDF errors from the same set (upper panel). Also shown is the ratio of the same PDF sets to the central value of CTEQ6.6 (lower panel).}
	\label{fig:intrinsic_bottom_fullpdf}
\end{figure}

In Fig.~\ref{fig:intrinsic_bottom_fullpdf} we show the sum of 
the intrinsic bottom PDF $b_1$ and the dynamically generated PDF $b_0$ from CTEQ6.6 for
different normalizations of the IB component, namely 0.01 and 0.035 $\times m_c^2/m_b^2$.
We compare this sum to the asymmetric uncertainties\footnote{The asymmetric errors are computed following \cite{Stump:2003yu,Nadolsky:2001yg}.} 
of the CTEQ6.6 PDF set (upper panel). 
In the same figure is also shown the ratio of the same PDFs to the central value of CTEQ6.6 (lower panel). 
As can be seen, the IB curve with the 0.035 $\times m_c^2/m_b^2$ normalization clearly lies outside the uncertainty band
whereas the one with the smaller normalization is marginally outside the band (up to $x\lesssim 0.6$). 

If we are looking for new physics with couplings proportional
to the mass, the suppression of IB compared to the IC would be partly compensated by the 
square of the coupling. For a more detailed study of the relevant parton-parton luminosities
please see Ref.\ \cite{Lyonnet:2015dca}.

\section{Collider observables}
\label{sec:observables}

Several collider observables receive large contributions from heavy quark initiated subprocesses
and are hence potentially sensitive to an intrinsic charm content in the nucleon.
In order to expect optimal effects the heavy quark PDF should be probed at large $x \gtrsim 0.2$ (for light-front models)
and not too large factorization scales. This kinematic region is best accessible at lower energies in the center-of-mass system (cms)
and/or large rapidities. Therefore, a fixed target experiment like AFTER@LHC 
\cite{Brodsky:2012vg,Lansberg:2012kf,Lansberg:2013wpx,Rakotozafindrabe:2013cmt}
operating at a cms energy $\sqrt{s}=115$ GeV with a high luminosity is ideally suited for searches of IC effects.
In the following we review some of the collider processes which have been studied in the literature in this respect.

\subsection{Open heavy flavor production}

Inclusive charm hadron ($D^0, D^+, D^{\star +}, \Lambda_c, \ldots$) production in hadronic collisions
was advocated in Ref.\ \cite{Kniehl:2009ar} as a laboratory to probe IC inside the colliding hadrons.
In this analysis, predictions for the differential cross section in dependence of the transverse momentum $p_T$
were obtained in the general-mass variable-flavor-number scheme (GM-VFNS) 
\cite{Kniehl:2004fy,Kniehl:2005mk,Kniehl:2005st}
at next-to-leading order (NLO).
In this scheme, the charm quark is an active parton and the differential cross sections of inclusive charm meson production 
depend heavily on the PDF of the charm quark.
The sensitivity of these cross sections to IC was studied for the Tevatron at a cms energy of 1960 GeV and the Relativistic Heavy Ion Collider (RHIC) 
at cms energies of 200 GeV (RHIC200) and 500 GeV (RHIC500). 
The different IC models from the CTEQ6.5c global analysis \cite{Pumplin:2007wg} were employed
together with the fragmentation functions for charm mesons from Ref.~\cite{Kneesch:2007ey}.
While the effects at the Tevatron were found to be very moderate and likely not testable, large enhancements were found at RHIC200
reaching values of $\sim 3$ at $p_T=20$ GeV. Unfortunately, the measurements at RHIC200 are limited by the luminosity.
At RHIC500 the cross section is increased by about a factor 3.6. However, the sensitivity to IC for the light-front models is greatly reduced.

More recently, the GM-VFNS was applied to obtain predictions for the production of inclusive $D$ mesons at the LHC for
a cms energy of 7 TeV (LHC7) \cite{Kniehl:2012ti}.
It was found that the production cross sections at large rapidities $y \gtrsim 4$ are sensitive to an IC component.
These predictions can be tested by measurements at forward rapidities with the LHCb detector.

\begin{figure*}[thp!]
\centering
\includegraphics[width=0.48\textwidth]{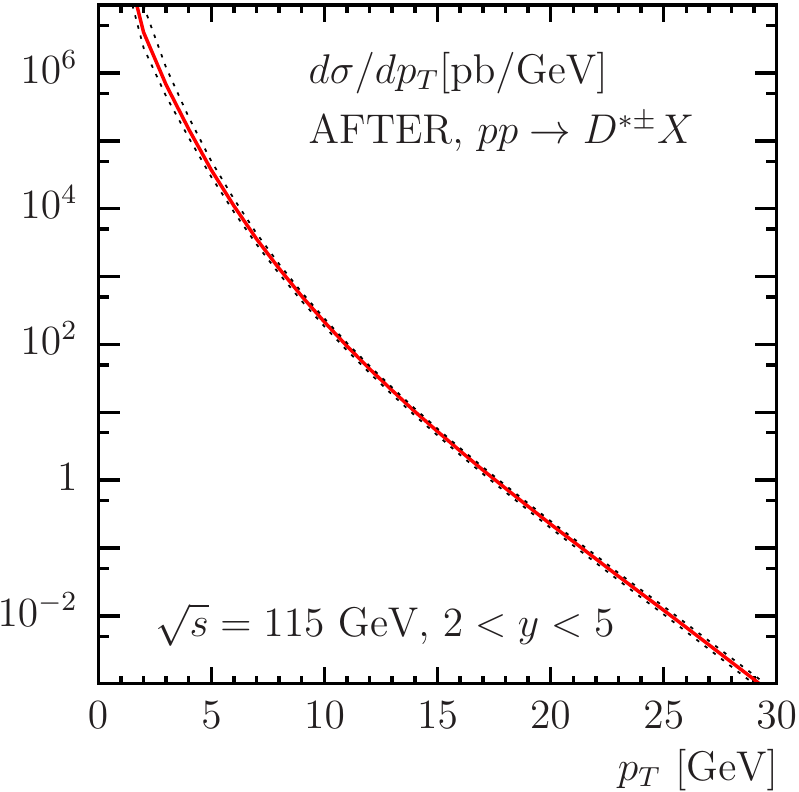}
\includegraphics[width=0.45\textwidth]{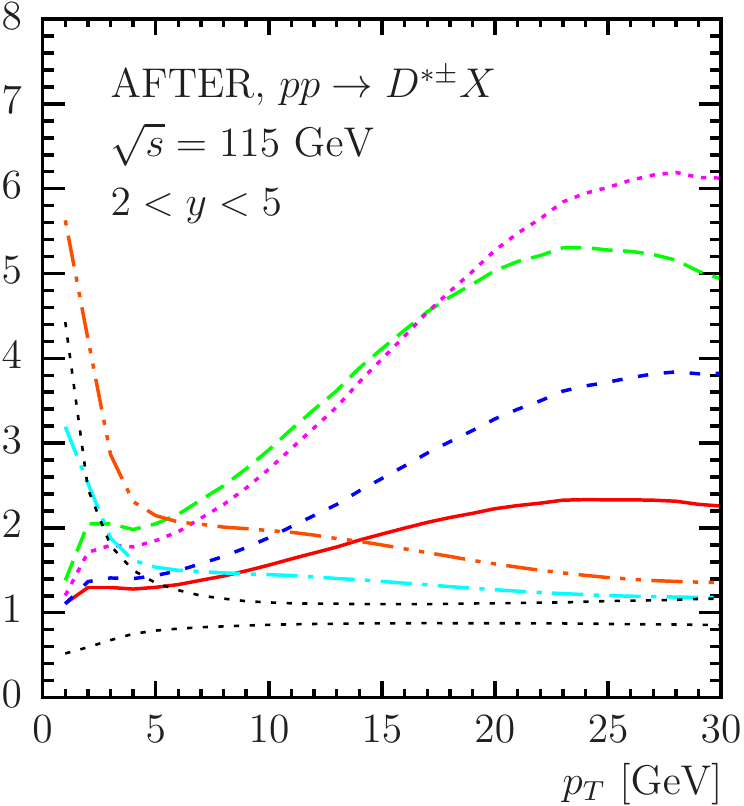}
\caption{ (Color online) 
NLO predictions for inclusive $D^\star$ meson production at AFTER@LHC vs the transverse
momentum of the $D$ meson. (Left) Differential cross section on an absolute scale without intrinsic charm.
(Right) Ratio w.r.t. to the central prediction of the left plot. Shown are results using the IC parametrizations
from Ref.\ \cite{Pumplin:2007wg} for $n=1$ (red, solid line), 2 (violet, dotted line), 3 (blue, dashed line), 4 (green, long dashed line),
5 (cyan, dot-dashed line), 6 (orange, double-dot-dashed line).
In both figures, the black dotted lines have been obtained by varying the renormalization scale around the central choice ($\mu_R = m_T$) 
to $\mu_R= m_T/2$ (upper line) and $\mu_R = 2 m_T$ (lower line). 
}
\label{fig:D+X} 
\end{figure*}

The ideal experiment to search for the effects of IC would be a high luminosity fixed target experiment
such as AFTER@LHC operating at a cms energy of 115 GeV.
In Fig.\ \ref{fig:D+X} we show results for inclusive $D^\star$ meson production as a function of the transverse momentum
of the $D^\star$ meson and integrated over the rapidity range $2 < y < 5$ (in the laboratory frame)
in essentially the same setup as in Ref.\ \cite{Kniehl:2009ar} to which we refer for details.
The only difference is that, following Ref.~\cite{Kniehl:2015fla}, the default choice for the renormalization
and factorization scales is $\mu_R = m_T$, $\mu_F = \mu_F' = m_T / 2$ where $m_T = \sqrt{p_T^2 + m^2}$ is the transverse mass.
The theoretical predictions are shown on an absolute scale in Fig.~\ref{fig:D+X} (left) and as
a ratio with respect to the default results in Fig.~\ref{fig:D+X} (right).
In both figures, the black dotted lines have been obtained by varying the renormalization scale around the central choice 
to $\mu_R= m_T/2$ (upper line) and $\mu_R = 2 m_T$ (lower line). 
In the right figure we repeat the calculation of the central prediction in turn with PDF sets 
CTEQ6.5Cn for $n=1,\ldots,6$ and normalize the outcome to the default prediction with zero IC of
Fig.~\ref{fig:D+X} (left). 
We observe that the ratios for $n=1,2,3,4$ corresponding to the BHPS ($n=1,2$) or meson-cloud ($n=3,4$) models 
become very large at large $p_T$. Indeed, the default cross section can be increased by more than a factor 5 at $p_T= 20$ GeV
in scenarios with maximally allowed intrinsic charm ($n=2,4$). Even for the IC sets with smaller normalization ($n=1,3$) corresponding to
$\langle x \rangle_{c_1 + \overline c_1} = 0.57\%$ and $\langle x \rangle_{c_1 + \overline c_1} = 0.96\%$ the cross section would be
enhanced by a factor larger than 2 (red solid line) or 3 (blue dashed line) at $p_T=20$ GeV.
It is also interesting to note that the phenomenological models for a sea like IC ($n=5,6$) lead to a significant enhancement
of the cross section at small $p_T \sim m_c$ which would be probed at AFTER@LHC as well.

\subsection{Production of a photon in association with a charm quark}

Another process with a wide range of phenomenological applications in $pp$, $pA$, and $AA$ collisions
\cite{Stavreva:2009vi,Stavreva:2010mw,Stavreva:2012aa} which is very sensitive to the heavy quark PDF
is the associated production of a photon with a heavy quark.
A dedicated study of this process at the LHC operating at $\sqrt{s}=8$ TeV (LHC8) was performed
in Refs.\ \cite{Bednyakov:2013zta,Bednyakov:2014pqa} where it was demonstrated that the existence of
IC in the proton can be visible at large transverse momenta of the photons and heavy quark jets at rapidities
$1.5 < |y_\gamma|<2.4, |y_c|<2.4$.
Indeed, for the BHPS model the cross section can be enhanced by a factor of 2-3 for $p_T^\gamma > 200$ GeV
(see Fig.\ 5 in  \cite{Bednyakov:2014pqa}).
This comes with the penalty  that the cross section falls rapidly with increasing transverse momentum so that this measurement
will be limited by statistics.

\begin{figure*}[thp!]
\centering
\includegraphics[width=0.48\textwidth]{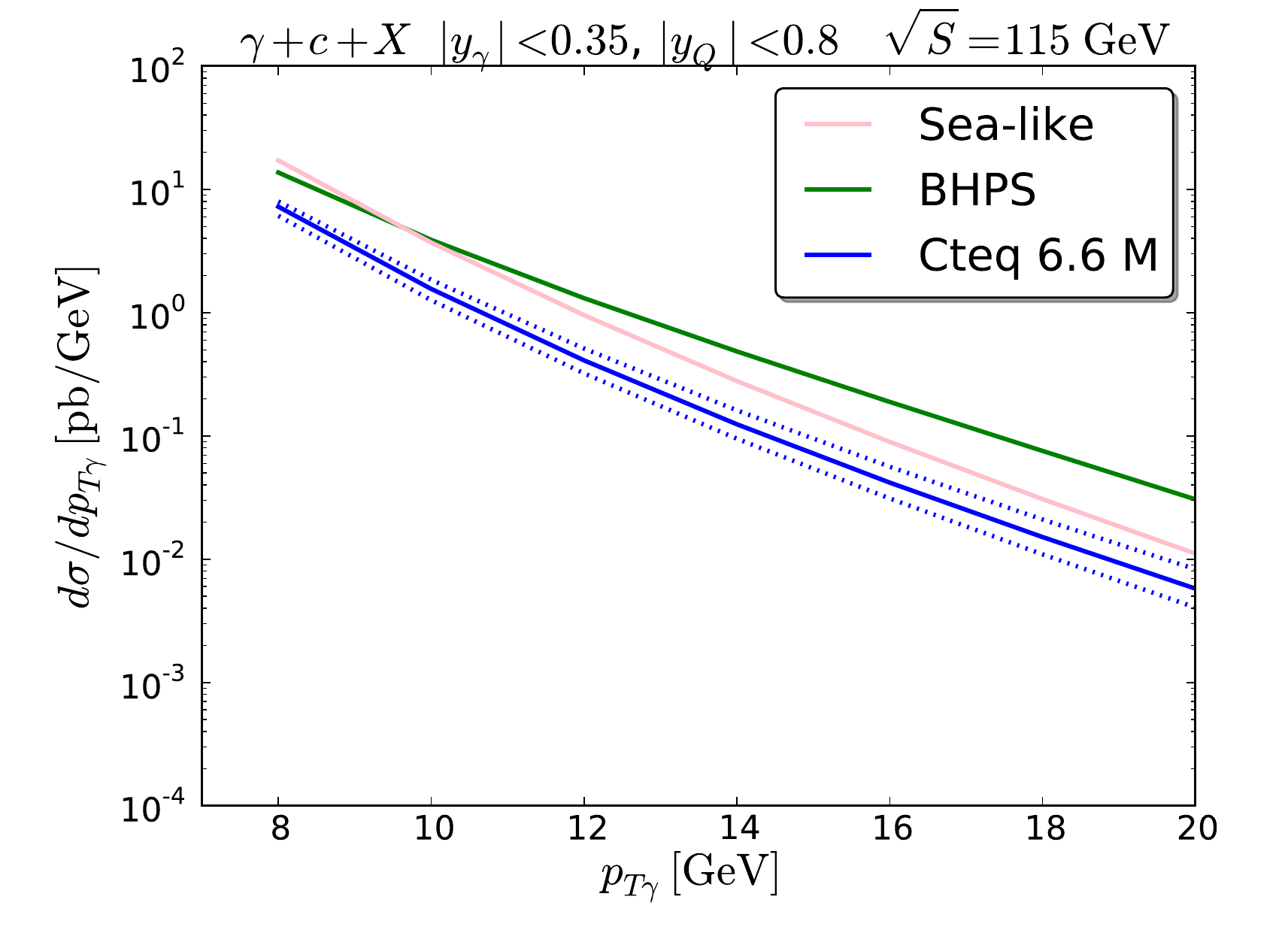}
\includegraphics[width=0.45\textwidth]{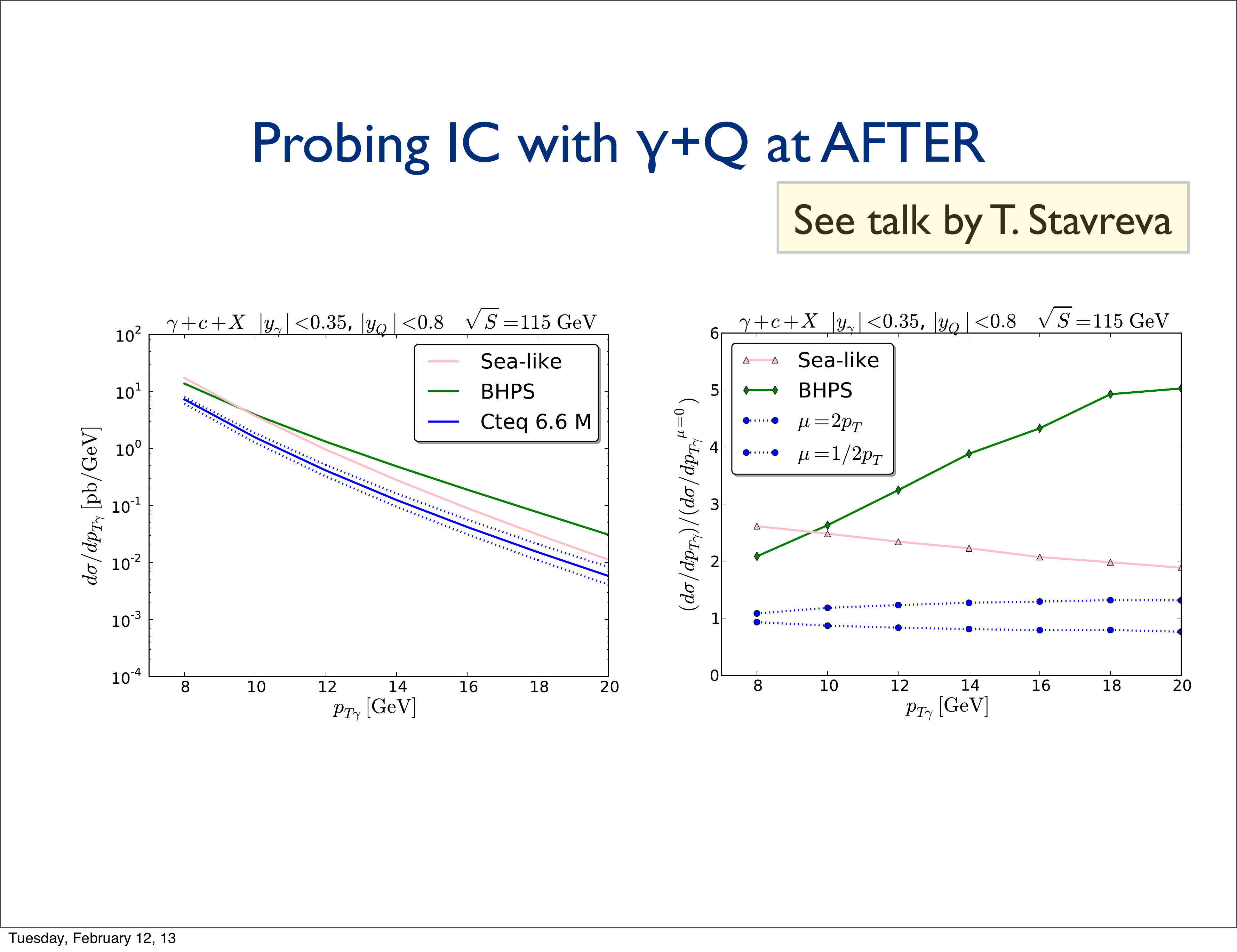}
\caption{ (Color online) 
NLO predictions for the production of a prompt photon in association with a
charm quark jet in $pp$ collisions at AFTER@LHC vs the transverse momentum of the photon.
Shown are results for an BHPS and a sea like intrinsic charm using the CTEQ6.6c PDFs.
For comparison, the predictions without an IC using the CTEQ6.6M PDFs are shown as well together
with the uncertainty band obtained by varying the central factorization scale $\mu_F=p_T^\gamma$ a factor
2 up and down (blue, dotted curves).
The right panel depicts the ratio of the curves in the left panel with respect to the central prediction without intrinsic charm.
}
\label{fig:gamma+c} 
\end{figure*}

Again, as for open heavy flavor production, the lower cms energy together with the high luminosity makes a fixed target experiment
like AFTER@LHC the ideal place to discover IC using $\gamma+c$ production.
This can be seen in Fig.\ \ref{fig:gamma+c}, where the differential cross section is enhanced by a factor 5 at $p_T^\gamma=20$ GeV (right panel)
with a not too small cross section (left panel).

\subsection{Vector boson production}

Dulat {\it et al.} \cite{Dulat:2013hea} studied the sensitivity of $W^\pm$
and $Z^0$ production to the presence of IC.  Vector boson production at the LHC
is an interesting ground for IC because it is at relatively large $x$ for 
colliders and $Z^0 \rightarrow l^+ l^-$ is a rather clean final state.
They did a NNLO calculation of
$W$ and $Z$ production including IC based on their global fits at 
$\sqrt{s} = 8$ and 14 TeV.  They also studied the ratio 
$d\sigma_{W^+ + W^-}(y)/d\sigma_{Z^0}(y)$ relative to the result with no IC.
Neither of these calculations showed an effect larger than the uncertainties
due to the CT10 sets themselves.  However, when the $Z^0$ $p_T$ distribution
with IC was compared to that without, they saw a factor of two enhancement at
$p_T \sim 500$ GeV for $\sqrt{s} = 8$ TeV in the range $|\eta| < 2.1$.  The
corresponding enhancement at 14 TeV was smaller at the same $p_T$ because the  
$x$ value reached is reduced at the higher energy \cite{Dulat:2013hea}.

\begin{figure*}[thp!]
\includegraphics[width=0.45\textwidth]{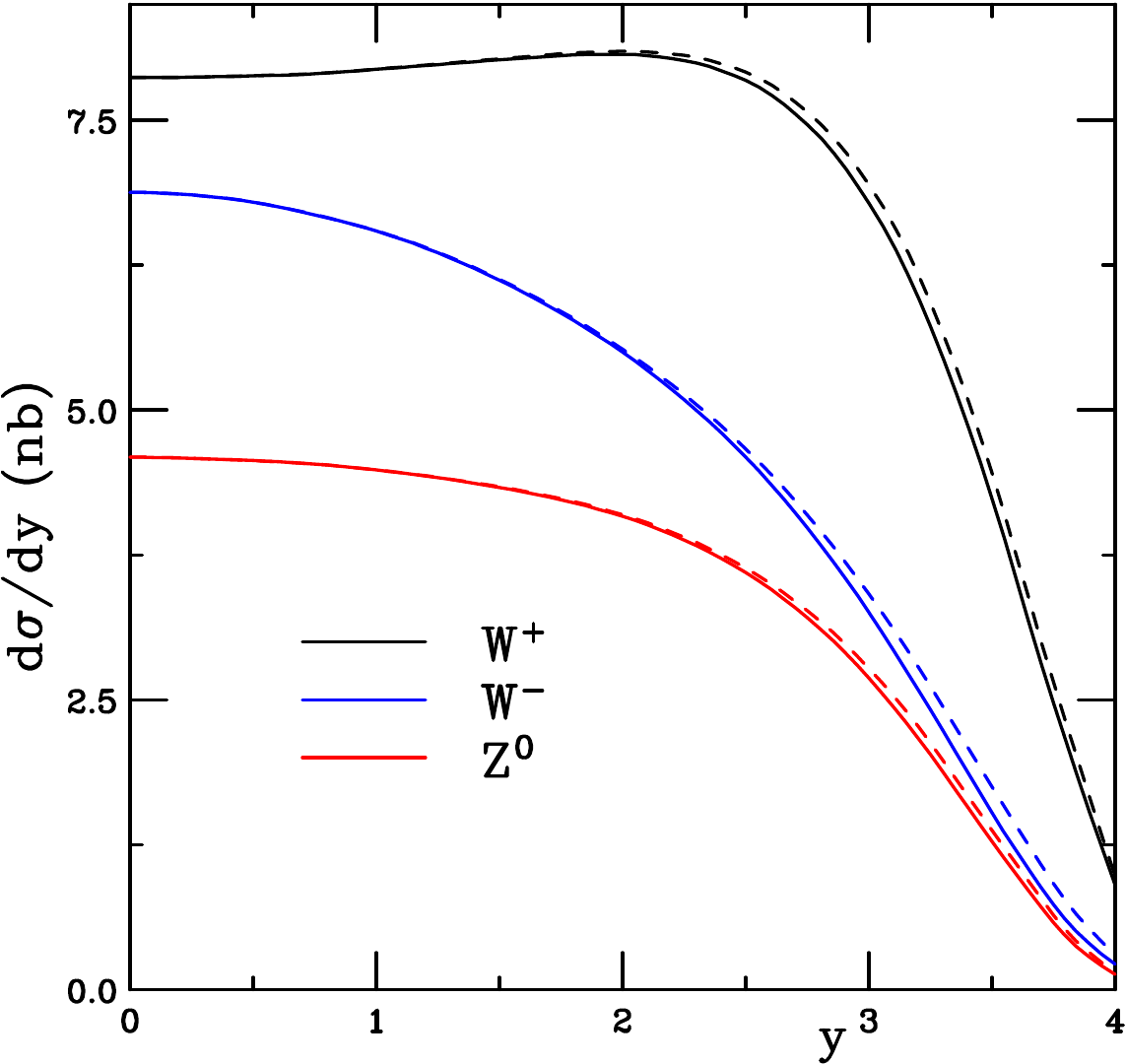}
\includegraphics[width=0.45\textwidth]{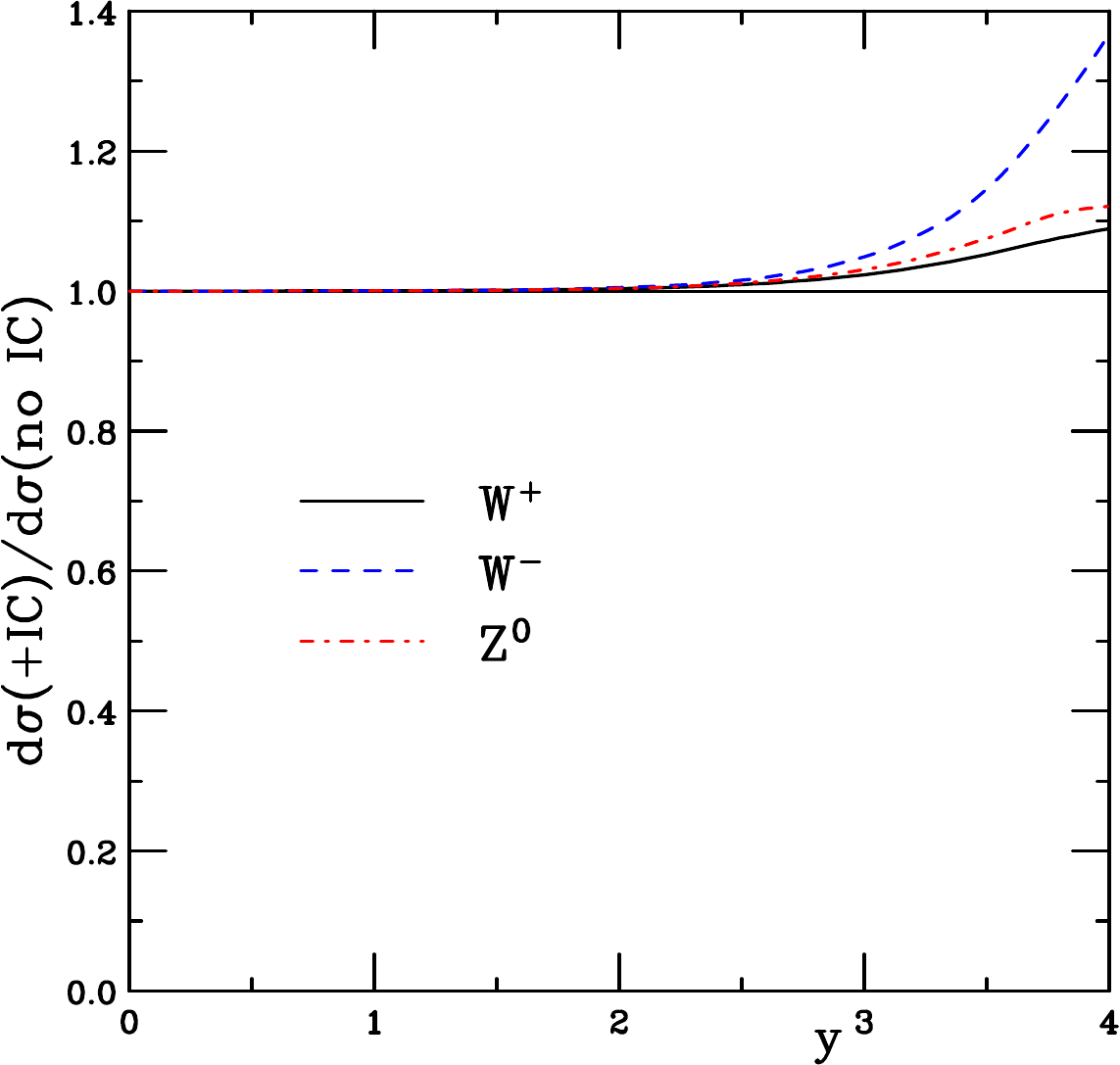}
\caption{(Color online) The $W^+$ (black), $W^-$ (blue) and $Z^0$ (red)
rapidity distributions (left).  The solid curves are the results without
IC while the dashed curves include 1\% BHPS IC.  The ratios of the dashed
curves to the solid curves, showing the enhancement of the rapidity 
distributions due to IC for $W^+$ (solid black), $W^-$ (blue dashed) and $Z^0$
(red dot dashed) are shown in the right plot.
}
\label{fig:RV_vecbos}
\end{figure*}

We show a simple test case here for $W$ and $Z$ production to NLO at
$\sqrt{s} = 7$ TeV.  We use only the BHPS IC parameterization for the 
five-particle Fock state, shown in Eq.~(\ref{eq:IC}).  We assume a 1\%
normalization and no $Q^2$ evolution to maximize the possible effect
at forward rapidity.  The $p_T$-integrated rapidity distribution is shown in
Fig~\ref{fig:RV_vecbos}, as is the ratio of the result with IC to that without
as a function of rapidity.  
The rapidity distributions without IC are given by the solid curves while
the dashed curves are the calculations with the BHPS IC contribution to the
charm parton density.  With BHPS IC, one expects enhancement only at
forward rapidity.  The enhancement from IC appears for $|y| > 2.5$.  Note
that if the sea-like IC would be used instead, the enhancement would be small
but finite over all rapidity.

The $W^+$ cross section is largest and most forward peaked, because of the
$u \overline d$ contribution.  The contribution from the $c \overline d$ part
is a very small addition since the $u$ valence contribution is large and peaks
at large $x$, making the $y$ distribution larger at $|y| \sim 2$ than at
$y = 0$.  Indeed, it gives the smallest IC contribution.
The $W^-$ distribution should have the largest possible contribution from IC
because both the $d \overline u$ and $d \overline c$ peak at low $x$ and
because the $d$ valence distribution peaks at lower $x$ so that the $W^-$
rapidity distribution has a maximum at $y=0$. At $|y| \sim 4$, the IC enhancement
is $\sim 40$\%.  Finally, the $Z^0$ distribution, with a plateau over 
$|y| < 1.5$, also has a very small IC contribution because the charm enhancement
only comes through $c \overline c$.

Such IC enhancements are only visible outside the midrapidity acceptance of
the collider detector coverage of CMS and ATLAS.  However, LHCb or ALICE
cover this forward rapidity range with muons and could detect forward $Z^0$.
They could also look at the lepton rapidity asymmetry, 
$(W^+ - W^-)/(W^+ + W^-)$, at forward rapidity.  The statistical accuracy
of the measurement would need to be high to distinguish an IC enhancement from
the no IC result, especially since the 1\% BHPS IC is likely an upper limit
on this enhancement.  Note that the higher energy 
of LHC Run 2 will reduce the potential enhancement even though it would increase
the rates.

\section{Conclusions}
\label{sec:conclusions}

The existence of non-perturbative intrinsic charm and bottom components is
a fundamental prediction of QCD. 
In this article, we have reviewed the current status of our understanding of 
this intrinsic heavy quark content of the nucleon which yet remains to be
confirmed experimentally.
In particular, after introducing theoretical models predicting the intrinsic heavy quark
distributions
we have turned to a summary of the available information on intrinsic charm coming from
global analyses of parton distribution functions.
There are no global analyses of intrinsic bottom available and we have described
how IB can be modeled in order to explore its impact on collider observables
keeping in mind that bottom quark initiated subprocesses play an important role
in certain electroweak observables and in models for physics beyond the Standard Model.
We then have turned to a discussion of collider processes where IC could be discovered.
Generally, the effects of IC are larger at colliders with a lower center-of-mass energy 
and for hard processes with moderate factorization scales.
Therefore, a high-luminosity fixed target experiment like AFTER@LHC operating at a center-of-mass energy
$\sqrt{s}=115$ GeV would be ideally suited to discover or constrain IC.

%%%%%%%%%%%%%%%%%%%%%%%%%%%%%%%%%%%%%%%%%%%%%%%%%%%%%%%%%%%%%%%%%%%%%%%

\section*{Acknowledgments}
We are grateful to T.~Stavreva for providing Fig.~\ref{fig:gamma+c}.
The work of S.~J.~B. was supported by the Department of Energy Contract No.~DE-AC02-76SF00515.
The work of RV was performed under the auspices of the U.S.\
Department of Energy by Lawrence Livermore National Laboratory under
Contract DE-AC52-07NA27344.

\bibliographystyle{apsrev4-1}
\bibliography{review_iQ}

\end{document}